\documentclass[prl,twocolumn,showpacs,floatfix]{revtex4-1}
\usepackage{graphics,epsfig,amsfonts,amssymb,amsmath}
\begin{document}

\title{Magnetoelectric coupling at the interface of BiFeO$_3$/La$_{0.7}$Sr$_{0.3}$MnO$_3$ multilayers}
\author{M.J. Calder\'on,$^1$ S. Liang,$^{2,3}$ R. Yu,$^4$ J. Salafranca,$^5$ S. Dong,$^{6}$ S. Yunoki,$^{7,8}$ \\ L. Brey,$^1$ A. Moreo,$^{2,3}$ and E. Dagotto$^{2,3}$}

\affiliation{
$^1$Instituto de Ciencia de Materiales de Madrid (CSIC), Cantoblanco, 28049 Madrid, Spain \\
$^2$Department of Physics and Astronomy, University of Tennessee, Knoxville, Tennessee 37996, USA \\
$^3$Materials Science and Technology Division, Oak Ridge National Lab, Oak Ridge, Tennessee 37831, USA\\
$^4$Department of Physics and Astronomy, Rice University, Houston, Texas 7705, USA \\
$^5$Departamento de F\'isica Aplicada III, Universidad Complutense de Madrid, 28040 Madrid, Spain \\
$^6$Department of Physics, Southeast University, Nanjing 211189, China \\
$^7$Computational Condensed Matter Physics Laboratory, RIKEN, Wako, Saitama 351-0198, Japan \\
$^8$CREST, Japan Science and Technology Agency (JST), Kawaguchi, Saitama 332-0012, Japan}
\date{\today}

\begin{abstract}
Electric-field controlled exchange bias in a heterostructure composed of the ferromagnetic manganite
La$_{0.7}$Sr$_{0.3}$MO$_3$ and the ferroelectric antiferromagnetic BiFeO$_3$ has recently been demonstrated experimentally. By means of a microscopic model Hamiltonian we provide a possible explanation of the origin of this magnetoelectric coupling. We find, in agreement with experimental results, a net ferromagnetic moment at the BiFeO$_3$ interface. The induced ferromagnetic moment is the result of the competition between the $e_{\rm g}$-electrons double exchange 
and the $t_{\rm 2g}$-spins antiferromagnetic superexchange that dominate in bulk BiFeO$_3$.
The balance of these simultaneous ferromagnetic and antiferromagnetic tendencies is strongly affected by the
interfacial electronic charge density which, in turn, can be 
controlled by the BiFeO$_3$ ferroelectric polarization.  
\end{abstract}

\maketitle

{\it Introduction.}
The quest for efficient electric field control of magnetic properties 
has encouraged research on materials with a strong coupling between the 
magnetic and dielectric degrees of freedom~\cite{mathurNat06}.  
Such a control would find applications in
magnetic field storage and sensors, constituting a major 
step forward in the field of spintronics. However, so far no bulk material 
seems to possess the required characteristics, including 
working near room temperature. For these reasons, the field effect device
presented in Ref.~\cite{BFO_LSMO_NM} signals a new route~\cite{review-multiferroics} to achieving those
goals by growing a few nanometers thick layer of La$_{0.7}$Sr$_{0.3}$MnO$_3$ (LSMO),
a ferromagnetic (FM) metal, on an antiferromagnetic (AFM) and ferroelectric (FE) material, 
BiFeO$_3$ (BFO)~\cite{BFO_LSMO_NM,BFO_LSMO_PRL}.  
These experiments provided evidence for an induced FM
moment in BFO at the interface. 
This magnetic moment is strongly affected by the BFO polarization, 
which results in an electric field control of the LSMO exchange bias (EB), 
and concomitant control of the LSMO magnetization. 

\begin{figure} 
\leavevmode
\includegraphics[clip,width=0.45\textwidth]{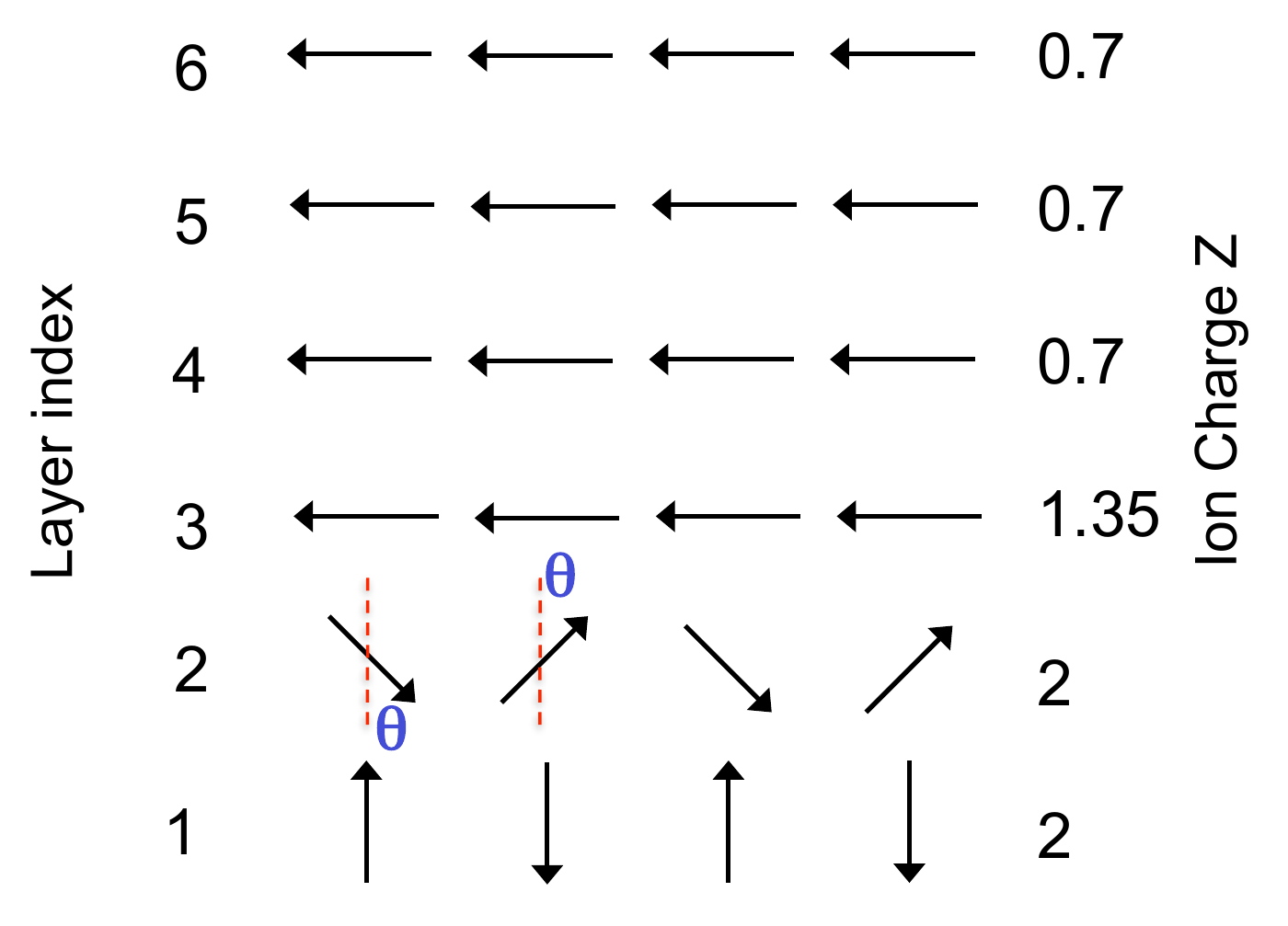}
\vskip -0.5cm 
\caption{(color online) Proposed spin order in the BFO-LSMO 
heterostructure (only half of the structure is shown, the other half is symmetric). 
The two first ($1,2$) and two last ($11,12$) planes have ionic charge Z=$+2$, 
and correspond to BFO. Layers $3$ and $10$ define the interface, with 
Z=$(0.7+2)/2$. The central planes ($4$ to $9$) have Z=$0.7$ and
correspond to LSMO. 
}
\label{spins}
\vskip -0.5cm
\end{figure}

Magnetic moments induced at the interface of perovskite-based oxide
materials have been previously reported. 
For example, a net magnetic moment was 
induced in an AFM manganite when grown in a multilayer
with a FM manganite~\cite{niebieskikwiat07,sefroui10}, at the interface of a superconducting cuprate with La$_{0.67}$Ca$_{0.33}$MO$_3$~\cite{Chakhalian}, and at LSMO/SrTiO$_3$ interfaces~\cite{Barriocanal_Ti}.
Electronic energy loss measurements indicate that a
charge redistribution takes place~\cite{Chakhalian,Barriocanal_Ti}. 
Theoretically, the origin of the induced magnetic moment
in~\cite{niebieskikwiat07} was explained in terms of charge transfer 
and a double exchange (DE) type
interaction~\cite{TMR}. Charge transfer, together with orbital reconstruction, is also believed to play a
role at cuprate/La$_{0.67}$Ca$_{0.33}$MO$_3$ interfaces~\cite{Chakhalian}. 
In \cite{BFO_LSMO_PRL,Okamoto_10} the BFO induced FM moment in LSMO/BFO heterostructures
was attributed to Fe-Mn hybridization, which is associated with charge transfer.  
The recently observed coupling of the EB with an induced
magnetization near the interface, simultaneously controlled by the FE
polarization~\cite{BFO_LSMO_NM,BFO_LSMO_PRL}, defines a new complex phenomenon that requires a better theoretical
understanding~\cite{Suhai_EB}. 

In this letter, a microscopic model for the transition metal $d$-electrons is shown to
explain the main properties of the BFO/LSMO interface. 
Within this model, the magnetic moments of the Fe ions develop
a net FM moment close to the interface as a consequence of both charge and
orbital redistribution, while they remain AFM ordered
far from the interface. Interestingly, the direction and
magnitude of the BFO moment with respect to the magnetization of the LSMO
layer depend on the charge density at the interface. 
Small changes in the charge density due to the switching of the FE
polarization in BFO produces large modifications in the direction of the magnetic moment
induced in the Fe ions. This leads to the experimentally observed EB~\cite{BFO_LSMO_NM}. Our scenario is qualitatively different from that proposed in~\cite{BFO_LSMO_PRL}, which is based on the first LSMO layer, and that of~\cite{Suhai_EB}, which requires a spin-orbit coupling.

{\it Model.} Fig.~\ref{spins} sketches the $4 \times 4 \times 12$ supercell used in our calculations. 
Bulk LSMO (BFO) is in a FM (G-type AFM) state. A possible magnetic reconstruction near the interface is also indicated. Departures from the cubic perovskite lattice, as induced by strain and ferroelastic effects, are ignored for simplicity. The $z$-axis size ($8$ layers for LSMO and $4$ for BFO) is selected such that bulk behavior is recovered at the center of the composing films. 
 
Magnetism in the heterostructure arises from the transition metal $d$ electrons.
In manganites, the three $t_{\rm 2g}$ electrons are localized and are approximated by a single classical spin. 
AFM superexchange with neighboring sites is introduced via a Heisenberg interaction~\cite{Dagotto-review}.
The quantum itinerant $e_{\rm g}$ electrons are described by the DE model in the infinite Hund's coupling limit.
Several aspects of the rich physics of bulk~\cite{Dagotto-review,MatToday,Sen_10,breyPRL04} and
heterostructured manganites~\cite{Brey_PS,TMR,dong08,rong09} have
been successfully addressed within this approach. However, model Hamiltonian approaches have not been used
before for BFO.

To properly consider the effects of charge leakage and orbital
hybridization at the interface~\cite{Okamoto_10}, the $e_{\rm g}$
electrons of {\it both} BFO and LSMO must be treated on equal footing. The model is:
\begin{eqnarray}\label{Ham}
\nonumber H &=& -\sum_{<i,j>,\alpha,\beta}
t_{ij} O_{i-j}^{\alpha\beta} \Omega_{ij} c_{i\alpha}^{\dagger}
c_{j \beta} + \sum_{<i,j>}
J^{\rm{AFM}}_{ij} \textbf{S}_{i} \cdot \textbf{S}_{{j}} \\
 & &
 + \sum_{i}(\tilde{\phi}_{i}+V_{i}-\mu) n_{i}.
\end{eqnarray}
Here, {\bf S}$_{i}$ represents the $t_{\rm 2g}$ spin at site
$i$, located either at the LSMO or the BFO side of the heterostructure, 
while $ J^{\rm{AFM}}_{ij} $ is the AFM superexchange parameter.
$c_{i\alpha}^{\dagger}$ creates an electron  
on an orbital centered at the transition metal site with $e_{\rm g}$ symmetry:
$\alpha$, $\beta$ = $|3z^2-r^2\rangle$, $|x^2-y^2\rangle$. 
The hopping term is modulated by the DE factor $\Omega_{ij}$~\cite{Dagotto-review} that depends on the
angle between {\bf S}$_{i}$ and {\bf S}$_{j}$ such that it is maximum for
 parallel alignment of spins at neighboring sites and
 zero for antiparallel alignment.  Hopping also depends 
on the overlap between the $\alpha$ and $\beta$ orbitals along the
direction $i-j$ through the
geometric factor $O^{\alpha\beta}_{i-j}$~\cite{Dagotto-review}.
In principle, the hopping parameters
$t_{ij}$ should depend on the material, and would be affected by lattice
distortions near the interface. We simplify the calculation by assuming a uniform hopping
parameter for the whole heterostructure ($t_{ij}$=$t$) ($t$ is the
energy unit). The superexchange coupling is more sensitive
to changes in lattice constants, hence
each material will be characterized by a different $J_{\rm AFM}$. 

In general, the DE term in Eq.(\ref{Ham}) favors FM
configurations that optimize the kinetic energy, while 
the superexchange term favors AFM phases. The third term contains the
different contributions to the site potential. Long range Coulomb interactions are
essential to control
charge transfer across the interface. This is included in the Hartree approximation
by setting 
\begin{equation}
\tilde{\phi}_{i} = \alpha \sum_{j\neq i} \frac{n_{j}-Z_j}{|\vec{r}_{i}-\vec{r}_{j}|} .\,
\end{equation}
The Coulomb interaction strength
is regulated by the parameter  $\alpha$, here assumed equal to $2t$~\cite{lin06,TMR}. 
For each $x$-$y$ plane, Z as illustrated in
Fig.~\ref{spins} is considered,  and 
the approximation is made that the background consists of point
charges that occupy the transition metal sites. Z=2 for BFO, while Z=0.7
for LSMO. An interfacial layer is considered with an intermediate value
of Z, to account for possible diffusions and different chemical
environments of the transition metal ions at the interface. 
$V_{i}$
includes the effect of the band offset 
between BFO and LSMO $V_{\rm offset}$, and the surface charge density due to
ferroelectricity $V_0$. The band offset is difficult to estimate, though 
expected to be small. From BFO electronic affinity~\cite{Yang10} and 
LSMO work functions~\cite{LSMO_wf}, we estimate $V_{\rm offset}=V_{\rm LSMO}-V_{\rm BFO} \sim 0.6t$. However, $V_{\rm offset}$ is here treated as an adjustable parameter: 
it is first set to zero and the results are then checked against
moderate changes in its value. The
chemical potential $\mu$ is chosen so that the overall system remains charge neutral.

\begin{figure}
\leavevmode
\includegraphics[clip,width=0.45\textwidth]{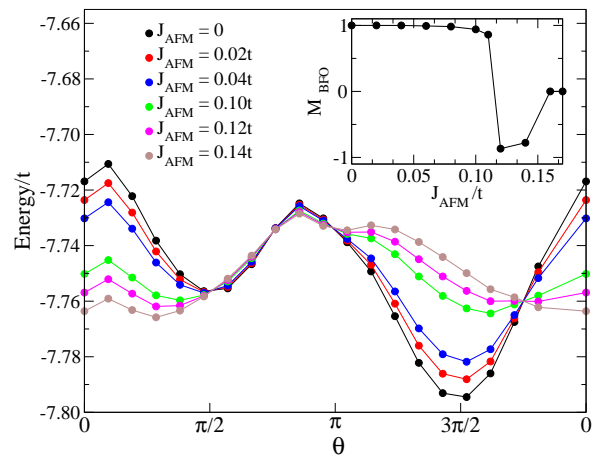}
\vskip -0.5cm
 \caption{(color online) (Main panel) Energy vs. canting
 angle $\theta$ for different values of $J_{\rm AFM}$. 
 (Inset) Value of the magnetization induced at the interfacial plane of
 BFO vs $J_{\rm AFM}$, as given by the position of the minima
 in the main panel. 
 }
\label{EvsTheta}
\vskip -0.5cm
\end{figure}

Several approximations are implicit in Eq.(\ref{Ham}). 
While the intraorbital Hubbard
interaction $U$ is effectively infinite due to the infinite 
Hund's coupling implicit in DE,
the interorbital $U^{\prime}$ might be important. However, it has been checked that
introducing a moderate $U^{\prime}$=2$t$ at the mean-field level does not
significantly affect the results. The coupling to 
Jahn-Teller lattice modes is not included
either, but it is widely accepted that they do not play an
important role in LSMO or BFO at any of the interfacial fillings of the $d$ bands. 
For each set of parameters and different $t_{\rm 2g}$ spin configurations,
Eq.(\ref{Ham}) was solved by
exact diagonalization. Periodic boundary conditions are used, with
a 5$\times$5$\times$1 mesh in reciprocal space.

{\it Results.} Our first important result is sketched in
Fig.~\ref{spins}. While LSMO remains FM, and most 
of BFO remains in the G-AFM state, a magnetic reconstruction  
takes place in the last atomic layer of BFO. The perpendicular orientation of the spins in bulk BFO (layer 1) relative to LSMO gives lower energy than a parallel orientation.
The magnetic state in the last BFO layer can be characterized by a single angle $\theta$, defined in
Fig.~\ref{spins}, which determines the magnetization of 
BFO close to the interface, $M_{\rm BFO}$.
For $\theta$=$\pi /2$, the last layer of BFO is FM but antiparallel to the LSMO magnetization, thus $M_{\rm BFO}$=$-1$; 
for $\theta$=$3 \pi /2$, the last layer of 
 BFO is  FM and  parallel to the LSMO magnetization, $M_{\rm BFO}$=$1$; and
 for $\theta$=$0, 2\pi$, there is no net magnetization in BFO. Regardless of the ground state value of
 $\theta$, the main magnetic reconstruction is confined to the BFO outmost
 layer, since spin canting  at the LSMO interfacial atomic layer is
 small. These results are remarkably independent of
details, such as whether there is an
interfacial layer with intermediate background charge. 
It also holds for several values of $J_{\rm AFM}$, as long as they
are reasonable (for the proper bulk phase diagram:
$J_{\rm AFM}^{\rm LSMO}<0.1$; $0.1<J_{\rm AFM}^{\rm BFO}<0.2$). $J_{\rm AFM}^{\rm Interface}$ (between layers 2 and 3) is expected to be some average of $J_{\rm AFM}^{\rm LSMO}$ and $J_{\rm AFM}^{\rm BFO}$. For simplicity, we use $J_{\rm AFM}$=$J_{\rm AFM}^{\rm Interface}$=$J_{\rm AFM}^{\rm BFO}$ and $J_{\rm AFM}^{\rm LSMO}=0$.

\begin{figure}
\leavevmode
\includegraphics[clip,width=0.4\textwidth]{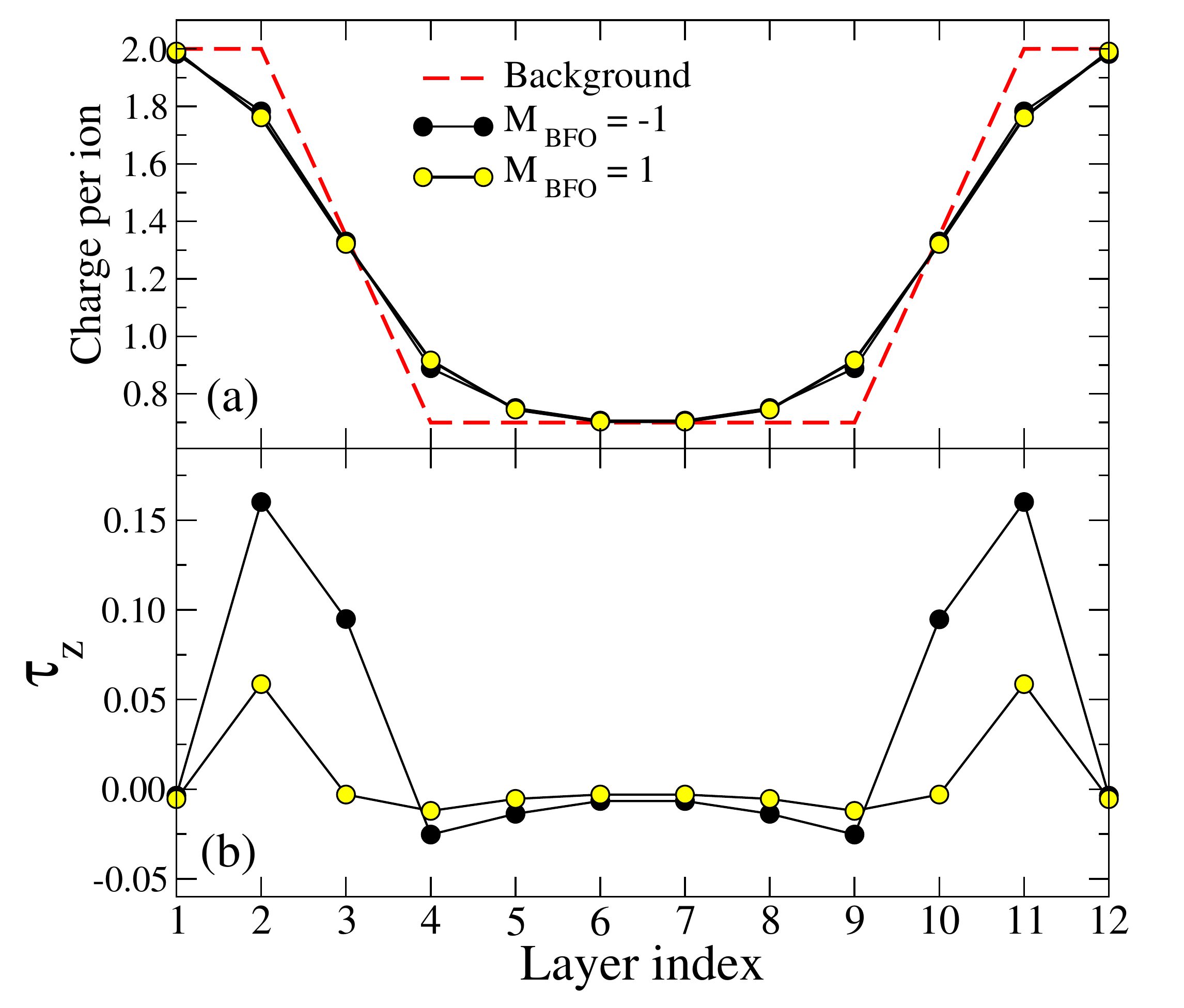}
\vskip -0.5cm
\caption{(color online) (a) Electronic charge for canting angles
 $\theta$$\sim$$\pi /2$ and 
$\theta$$\sim$$3 \pi /2$ (the dashed line is the positive charge background).
The charge density distribution depends only slightly on
 $\theta$. 
 (b) Orbital order described by
 the expectation value of 
 $\langle\tau _z \rangle$=$ n_{x^2 - y ^2}$-$n_{3z^2 -r^2}$. The $e_{\rm g}$ orbitals are equally populated
 ($\langle \tau_z \rangle$=0) at the BFO and LSMO bulk layers. 
} 
  \label{Density_Profile}
\vskip -0.5cm
\end{figure}

Fig.~\ref{EvsTheta} illustrates the energy as a function of
$\theta$ for different values of $J_{\rm AFM}$.  At small $J_{\rm AFM}$, the
energy is minimized at $\theta$=$3 \pi/2$: the BFO interfacial layer is FM and parallel
to the metallic LSMO and DE dominates. As $J_{\rm AFM}$ increases, the
minimum near $3 \pi /2$ moves  toward $2 \pi$. 
In addition, the minimum near $\pi /2$ decreases in energy and eventually has the lowest energy.
In this case the last layer of BFO is FM and antiparallel to the LSMO magnetization. This trends are summarized in the inset of Fig.~\ref{EvsTheta} where the magnetization induced in the last layer of BFO is plotted as a function of $J_{\rm AFM}$. The magnetization rotation (from
 $\theta$=$3 \pi/2$ to $\pi/2$) occurs at  $J_{\rm AFM}$$\sim$$0.12 t$
 and there is some small canting for $0.07t \lesssim J_{\rm AFM} \lesssim
 0.11 t$.  For $J_{\rm AFM} \gtrsim 0.16t$, the
 AFM coupling  is stronger than the DE and
 the canting angle is zero, namely, the interfacial BFO layer is
 AFM as in bulk. Several results obtained in the simplified discussion
presented here, such as Fig.~\ref{EvsTheta} and others based on the simple
assumption that the interfacial behavior is characterized by a single angle $\theta$, 
were also
confirmed numerically using 4$\times$4$\times$8 clusters, the Poisson equation, and
a minimization algorithm for the $t_{\rm 2g}$ classical spins~\cite{Broyden}.  

The observed charge redistribution and orbital reconstruction confirm the
importance of the DE mechanism.  Fig.~\ref{Density_Profile} shows the
charge $n_{i}$ and orbital polarization $\langle\tau _z \rangle= n_{x^2
- y ^2}- n_{3z^2 -r^2}$ for two values of $\theta$.    
The $e_{\rm g}$-charge profile is mainly determined by the background
charge except at the interface planes where there is a charge redistribution whose extension is controlled by the parameter $\alpha$~\cite{Brey_PS}: holes at 
the last atomic plane of BFO and extra electrons at LSMO favor a kinetic energy gain.
For $M_{\rm BFO}$=-1, the hopping between 
LSMO and the interfacial BFO layer is suppressed. However, even in this
case the Fe $d$-orbitals are not completely filled, and some
kinetic energy gain occurs, mainly within the $x$-$y$ plane. This asymmetry is
evidenced by a non-zero orbital polarization $\langle\tau_{z} \rangle$, at both sides of the interface (layers 2,3,10,11).
Since there is no electron-lattice coupling
here, the orbital polarization necessarily arises from
asymmetries in the orbital filling due to kinetic
energy gain. For $M_{\rm BFO}$=1
there is hopping across  the interface and a significant orbital polarization 
appears only at the BFO side (layers 2,11). The positive orbital polarization is caused by the suppression of the hopping in the $z$-direction due to the filled bands of bulk BFO leading to an enhancement of the hopping in the $x$-$y$ plane.  
A larger kinetic energy in this case compensates for the Coulomb energy
cost of an $e_{\rm g}$ density that further deviates 
from the background value. Although the difference in charge profile for
the two angles is small, it plays an important role in the phenomena
discussed next.

\begin{figure*}
\leavevmode
\includegraphics[clip,width=0.8\textwidth]{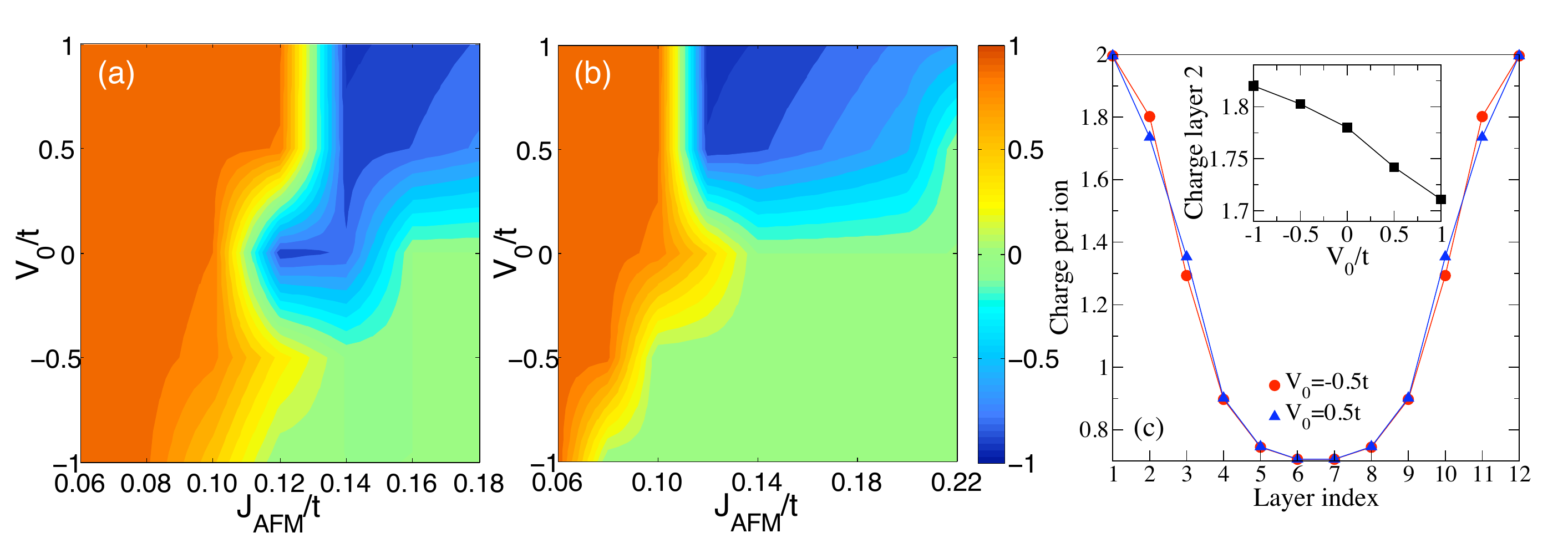}
\vskip -0.5cm
\caption{(color online) Magnetization induced in BFO ($M_{\rm BFO}$) as a function of $J_{\rm AFM}$ 
and  $V_0$ for (a)$V_{\rm offset}=0$ and (b)$V_{\rm offset}=0.6t$. (c) Electronic charge in the heterostructure for $V_0=-0.5t$ and $V_0=0.5t$. Inset: charge at layer 2 for $-1t \leq V_0 \leq 1t$.}
\label{bias_JAF}
\vskip -0.5cm
\end{figure*}

Consider now the effect of the BFO FE polarization. 
In our study, we assume that the main effect of switching the FE polarization is to
modify the induced charge density at the FE surface. 
As a consequence, the FE nature of BFO makes the heterostructure work as
a field-effect device~\cite{BFO_LSMO_NM}. The
direction of the FE polarization is modeled 
by introducing an additional potential $V_0$ at the surface of  
BFO (layers 2 and 11) which attracts ($V_0<0$) or repels ($V_0>0$) the charge [see Fig.~\ref{bias_JAF} (c)]. 
Fig. \ref{bias_JAF}, our main result, shows the value of the induced magnetization in
BFO, $M_{\rm BFO}$, as a function of $J_{\rm AFM}$ and $V_0$, for 
(a) zero band-offset, and (b) $V_{\rm offset}$=$0.6t$. For small values of
$J_{\rm AFM}$, a FM BFO layer appears at the interface, parallel to the
magnetization on LSMO ($\theta$=$3\pi/2$). For larger values of $J_{\rm AFM}$
and an attractive $V_0$, the AFM solution ($\theta$=$0$) is
obtained. This is due to the fact that increasing the density of charge
towards $2$ produces a decrease in kinetic energy, so the gain in
superexchange energy dominates. For large values of $J_{\rm AFM}$ and
repulsive $V_0$, a FM BFO layer is obtained which is
antiparallel to the magnetization on LSMO ($\theta$=$\pi/2$). This is
due to the decrease of the charge density (away from $2$) at the
interface which produces an increase of the $x$-$y$ plane kinetic energy, 
while the superexchange term gains energy by making 
the spins of layers $2$ and $3$ antiparallel. Equivalent results are found for different values of the band-offset. Therefore, 
Fig.~\ref{bias_JAF} explains the experimentally demonstrated control of the EB. 
The LSMO EB is determined by the magnetic order of the
last layer of BFO, via a partial pinning to the AF BFO bulk~\cite{Ohldag}, but this order is strongly affected by
an electric field (through changes in $V_0$). A magnetic field can also influence the EB,
by reducing the effective value of $J_{\rm AFM}$ and favoring FM
order in the last layer of BFO. 

{\it Conclusions.} A microscopic model that explains the
recently unveiled properties of the BFO/LSMO interface is proposed. 
The charges and spins couple via the DE and superexchange mechanisms, and our 
calculations show that the induced magnetic moment in BFO 
arises from charge transfer between the two materials. The spin arrangement generated at the BFO interfacial layer arises from 
the frustrating effect caused by the two competing (FM and AFM) tendencies in
adjacent layers with different electronic densities~\cite{rong09}. 
Our main result is that changing the sign of the 
BFO ferroelectric polarization modifies the extra charge near the interface, which in turn
strongly affects the magnitude and 
direction of the magnetic moment. This gives rise to the experimentally
observed magnetoelectric coupling, and clarifies the origin of the
recently observed electric field controlled exchange bias in LSMO/BFO heterostructures. 

M.J.C. and L.B. acknowledge funding from MICINN (Spain) 
through Grant No. FIS2009-08744. 
M.J.C. also acknowledges the Ram\'on y Cajal program, MICINN (Spain).
J.S. acknowledges European research Council Starting investigator Award STEMOX 239739. 
S.L., A.M., and E.D. are supported by the U.S. Department of Energy, 
Office of Basic Energy Sciences, Materials Sciences and Engineering Division.

\end{document}